\documentclass{elsart}
\usepackage{amssymb}

\newcommand{\tr}{\mathrm{Tr}}

\begin{document}

\begin{frontmatter}
\title{Decay rate and decoherence control in coupled dissipative cavities}

\author[UFMG]{A.~R. {Bosco de Magalh\~aes}}
\author[USP,PUC]{S.~G. Mokarzel}
\author[UFMG]{M.~C. Nemes}
\author[UFMGMAT,UFMG]{M.~O. {Terra Cunha}}

\address[UFMG]{Departamento de F\'\i sica, Universidade Federal de Minas
Gerais, C.P. 702, Belo Horizonte, 30123-970, Brazil}
\address[UFMGMAT]{Departamento de Matem\'atica, Universidade Federal de Minas Gerais, C.P. 702, Belo Horizonte,
30123-970, Brazil}
\address[USP]{Departamento de F\'{\i}sica--Matem\'atica, Instituto de
 F\'\i sica, Universidade de S\~ao Paulo,  C.P. 66318,  S\~ao Paulo, 05315-970, Brazil}
\address[PUC]{Departamento de F\'{\i}sica, Pontif\'{\i}cia Universidade 
Cat\'olica de S\~ao Paulo, R. Marqu\^es de Paranagu\'a, 111, S\~ao Paulo,
Brazil}

\begin{abstract}
We give a detailed account of the derivation of a master equation for two
coupled cavities in the presence of dissipation. The analytical solution is
presented and physical limits of interest are discussed. Firstly we show
that the decay rate of initial coherent states can be significantly modified
if the two cavities have different decay rates and are weakly coupled
through a wire. Moreover, we show that also decoherence rates can be
substantially altered by manipulation of physical parameters. Conditions for
experimental realizations are discussed.
\end{abstract}


\end{frontmatter}

\section{Introduction}

The past decade can celebrate having overcome several barriers, the most
important of them being the objections commonly levelled against the idea of
the observability of large scale quantum phenomena. The amazing
technological developments in several areas of physics allowed for the
experimental verification of essentially quantum effects like tunneling,
various kinds of interference and entanglement phenomena in mesoscopic
physics. So we have reached the point where we need to know rather urgently
what are the mechanisms which cause decoherence, particularly at the
mesoscopic scale, since controlling such mechanism has immediate and
profound technological consequences, besides its value for the fundamentals
of Quantum Mechanics and the (unsolved) problem of the classical limit.

Modern cavity quantum electrodynamics (cavity QED) sheds light onto the most
fundamental aspects of coherence and decoherence in quantum mechanics. One
of the main advantages of such experiments is that they can be described by
elementary models, being at the same time rich enough to reveal intriguing
subtleties of the interplay of coherent dynamics with external coupling. It
represents therefore a unique paradigm for matching theory with experiment
in the study of quantum coherence. In particular a very important experiment
has been proposed \cite{Davidovich1} and realized \cite{Brune1} where one
monitors the coherence loss of a superposition state.

The predominant component of decoherence in cavity QED systems corresponds
simply to the scape or emission of photons, either by absorption into the
cavity walls or by scattering or transmission into electromagnetic modes
outside the cavity. This decoherence mechanism fits well into the scheme of
coupling to extended modes. The description of decoherence through a linear
coupling to a set of harmonic oscillators seems to work well, both
qualitatively and quantitatively.

The purpose of the present work is to investigate dissipation and
decoherence effects on two field modes by generalizing the usual dissipation
mechanism for the present situation. The results can be experimentally
implemented in two distinct ways: in a regime with two modes of one cavity
close to resonance with an atomic transition, a task already implemented 
\cite{Rauschenbeutel1}, and in an experiment involving two cavities, as the
one proposed in ref. \cite{Davidovich2}, or in the teleportation scheme \cite
{Davidovich3}.

Our results show interesting physical effects for the situation when the
coupling of each mode with the reservoir is not independent. Despite of
being of difficult experimental implementation, we show which are the key
parameters or combinations of parameters which can lead to a decoherence
free subspace, or, more modestly, to decoherence control.

Moreover we investigate the coupling of two cavities through a wire and show
how dissipation can be controlled if the cavities are initially fed with
coherent states.

The counterintuitive effects here presented are consequences of quantum
destructive interference of coupling constants, a well studied theoretical
prevision \cite{Agarwal1} with interesting consequences as
electromagnetically induced transparency (EIT) \cite{Boller1}, lasing without
inversion \cite{Harris1} and narrowing of spectral lines \cite{Zhou1,Keitel1}.

The paper is organized as follows. In section II we derive the master
equation from a Hamiltonian for two coupled oscillators in the presence of
an environment, under some dynamical hypothesis. Section III presents the
exact formal solution of the Liouvillian by means of Lie-algebraic
techniques. There we discuss conditions for dissipation control. In section
IV we analyse the case with initially coherent states in the cavities and
present an interesting effect in the case of weak coupling between them.
Section V discusses the conditions for decoherence control through the study
of an initially pure entangled state. In section VI we comment on the
(difficult) experimental realization of the conditions found in the previous
sections.

\section{Master equation derivation}

We consider two modes linearly coupled both to each other and to the
environment. In both cases applying the Rotating Wave Approximation (RWA),
what is not too restrictive in the microwave domain \cite{Haroche1}. This
corresponds to the Hamiltonian
\begin{eqnarray}
\mathbf{H}_{o} &=&\mathbf{H}_{S}\mathbf{+H}_{R},  \nonumber \\
\mathbf{H}_{S} &=&\hbar \omega _{a}\mathbf{a}^{\dagger }\mathbf{a}+\hbar
\omega _{b}\mathbf{b}^{\dagger }\mathbf{b}+\hbar g\left( \mathbf{a}^{\dagger
}\mathbf{b}+\mathbf{b}^{\dagger }\mathbf{a}\right) ,  \nonumber \\
\mathbf{H}_{E} &=&\sum_{k}\hbar \omega _{k}\mathbf{c}_{k}^{\dagger }\mathbf{c%
}_{k},  \nonumber \\
\mathbf{H}_{int} &=&\sum_{k}\hbar \left( \alpha _{k}\mathbf{c}_{k}^{\dagger }%
\mathbf{a}+\alpha _{k}^{\ast }\mathbf{a}^{\dagger }\mathbf{c}_{k}\right)
+\sum_{k}\hbar \left( \beta _{k}\mathbf{c}_{k}^{\dagger }\mathbf{b}+\beta
_{k}^{\ast }\mathbf{b}^{\dagger }\mathbf{c}_{k}\right) ,
\label{HS}
\end{eqnarray}
where the operators $\mathbf{a}$ ($\mathbf{b}$) and $\mathbf{a}^{\dagger }$ (%
$\mathbf{b}^{\dagger }$) are annihilation and creation operators of the
field mode $A$ ($B$), directly coupled by the last term in eq.\ (\ref{HS}),
where $g$ is a real positive coupling constant. The environment is modelled,
as usual, by a set of harmonic oscillators with creation and annihilation
operators $\mathbf{c}_{k}^{\dagger }$ and $\mathbf{c}_{k}$, with complex
coupling constants $\alpha _{k}$ ($\beta _{k}$) to the mode $a$ ($b$). Much
of the results to be presented will depend on the existence of correlations
between $\alpha _{k}$ and $\beta _{k}$ for fixed $k$, as must become clear
later on.

It is convenient to work with the normal modes of $\mathbf{H}_{S}$:
\begin{eqnarray}
\mathbf{H}_{S} &=&\hbar \omega _{1}\mathbf{a}_{1}^{\dagger }\mathbf{a}%
_{1}+\hbar \omega _{2}\mathbf{a}_{2}^{\dagger }\mathbf{a}_{2},  \nonumber \\
H_{int} &=&\sum_{i=1}^{2}\sum_{k}\hbar \left( \alpha _{ik}\mathbf{c}%
_{k}^{\dagger }\mathbf{a}_{i}+\alpha _{ik}^{\ast }\mathbf{a}_{i}^{\dagger }
\mathbf{c}_{k}\right) ,
\end{eqnarray}
where
\begin{eqnarray}
\omega _{1} &=&\frac{\left( \omega _{a}+\omega _{b}\right) +\sqrt{\left(
\omega _{a}-\omega _{b}\right) ^{2}+4g^{2}}}{2},{ \ \ \ \ }\omega _{2}=%
\frac{\left( \omega _{a}+\omega _{b}\right) -\sqrt{\left( \omega _{a}-\omega
_{b}\right) ^{2}+4g^{2}}}{2},  \nonumber \\
\alpha _{1k} &=&\alpha _{k}\cos \theta +\beta _{k}\sin \theta ,{ \ \ \
\ }\alpha _{2k}=-\alpha _{k}\sin \theta +\beta _{k}\cos \theta ,  \nonumber
\\
\mathbf{a}_{1} &=&\mathbf{a}\cos \theta +\mathbf{b}\sin \theta ,{ \ \ \
\ }\mathbf{a}_{2}=-\mathbf{a}\sin \theta +\mathbf{b}\cos \theta ,  \nonumber
\\
\cos \theta  &=&\sqrt{\frac{1}{2}\left( 1+\frac{\omega _{a}-\omega _{b}}{%
\sqrt{\left( \omega _{a}-\omega _{b}\right) ^{2}+4g^{2}}}\right) },{ \
\ \ \ }\sin \theta =\sqrt{\frac{1}{2}\left( 1-\frac{\omega _{a}-\omega _{b}}{%
\sqrt{\left( \omega _{a}-\omega _{b}\right) ^{2}+4g^{2}}}\right) }.
\label{transf}
\end{eqnarray}
The operators $\mathbf{a}_{i}$ and $\mathbf{a}_{i}^{\dagger }$ obey the
usual commutation relations for bosons.

Time evolution of the complete density operator $\mathbf{\rho }$ is given by
von Neumann equation
\begin{equation}
\frac{d}{dt}\mathbf{\tilde{\rho}}\left( t\right) =\frac{-i}{\hbar }\left[
\tilde{\mathbf{H}}_{int}\left( t\right) ,\mathbf{\tilde{\rho}}\left(
t\right) \right] ,  \label{vonNewman}
\end{equation}
where the tilde over the operators tells us that they are written in the
interaction picture,
\begin{equation}
\tilde{\mathbf{O}}=\exp {\left\{ \frac{i}{\hbar }\mathbf{H}_{o}t\right\} }%
\mathbf{O}\exp {\left\{ \frac{-i}{\hbar }\mathbf{H}_{o}t\right\} }.
\end{equation}
Integrating eq. (\ref{vonNewman}) and iterating twice leads to
\begin{equation}
\mathbf{\tilde{\rho}}\left( t\right) -\mathbf{\tilde{\rho}}\left( 0\right) =-%
\frac{i}{\hbar }{\int_{0}^{t}}dt^{\prime }[\mathbf{\tilde{H}}_{int}\left(
t^{\prime }\right) ,\mathbf{\tilde{\rho}}\left( 0\right) ]-\frac{1}{\hbar
^{2}}{\int_{0}^{t}}dt^{\prime }{\int_{0}^{t^{\prime }}}dt^{\prime \prime }[%
\mathbf{\tilde{H}}_{int}\left( t^{\prime }\right) ,[\mathbf{\tilde{H}}%
_{int}\left( t^{\prime \prime }\right) ,\mathbf{\tilde{\rho}}\left( 0\right)
]]+\mathcal{O}^{3}\left( \mathbf{\tilde{H}}_{int}\right) .  \label{integr}
\end{equation}
Since the high quality factor of the cavity(ies), we consider weak coupling
of modes $A$ and $B$ to the environment, which allow us to disregard the
terms of third order in $\mathbf{\tilde{H}}_{int}$, denoted $\mathcal{O}%
^{3}\left( \mathbf{\tilde{H}}_{int}\right) $, in the following calculation.

Taking the trace over the environment degrees of freedom gives the
corresponding equation for the reduced density operator for the subsystem
composed by the two relevant modes (this subsystem will be called \textit{
system}) and eq. (\ref{integr}) then implies
\begin{eqnarray}
\mathbf{\tilde{\rho}}_{S}\left( t\right) -\mathbf{\tilde{\rho}}_{S}\left(
0\right)  &=&-i{\int_{0}^{t}}dt^{\prime }{\tr}_{E}{\sum_{i=1}^{2}}\left[
\mathbf{\Gamma }_{i}^{\dagger }\left( t^{\prime }\right) \mathbf{\tilde{a}}%
_{i}\left( t^{\prime }\right) +\mathbf{\Gamma }_{i}\left( t^{\prime }\right)
\mathbf{\tilde{a}}_{i}^{\mathbf{\dagger }}\left( t^{\prime }\right) ,\mathbf{%
\tilde{\rho}}\left( 0\right) \right]   \nonumber \\
&&-{\int_{0}^{t}}dt^{\prime }{\int_{0}^{\ t\prime }}dt^{\prime \prime }{\tr}_{E}\left\{ \sum_{i,j=1}^{2}
\begin{array}{c}
+\left(
\begin{array}{c}
\mathbf{\Gamma }_{i}^{\dagger }\left( t^{\prime }\right) \mathbf{\Gamma }%
_{j}^{\dagger }\left( t^{\prime \prime }\right) \mathbf{\tilde{a}}_{i}\left(
t^{\prime }\right) \mathbf{\tilde{a}}_{j}\left( t^{\prime \prime }\right)
\mathbf{\tilde{\rho}}\left( 0\right) +{\mathrm{h.c}}. \\
+\mathbf{\Gamma }_{i}\left( t^{\prime }\right) \mathbf{\Gamma }_{j}^{\dagger
}\left( t^{\prime \prime }\right) \mathbf{\tilde{a}}_{i}^{\mathbf{\dagger }%
}\left( t^{\prime }\right) \mathbf{\tilde{a}}_{j}\left( t^{\prime \prime
}\right) \mathbf{\tilde{\rho}}\left( 0\right) +{\mathrm{h.c.}}
\end{array}
\right)  \\
+\left(
\begin{array}{c}
\mathbf{\Gamma }_{i}^{\dagger }\left( t^{\prime }\right) \mathbf{\Gamma }%
_{j}\left( t^{\prime \prime }\right) \mathbf{\tilde{a}}_{i}\left( t^{\prime
}\right) \mathbf{\tilde{a}}_{j}^{\mathbf{\dagger }}\left( t^{\prime \prime
}\right) \mathbf{\tilde{\rho}}\left( 0\right) +{\mathrm{h.c.}} \\
+\mathbf{\Gamma }_{i}\left( t^{\prime }\right) \mathbf{\Gamma }_{j}\left(
t^{\prime \prime }\right) \mathbf{\tilde{a}}_{i}^{\mathbf{\dagger }}\left(
t^{\prime }\right) \mathbf{\tilde{a}}_{j}^{\mathbf{\dagger }}\left(
t^{\prime \prime }\right) \mathbf{\tilde{\rho}}\left( 0\right) +{\mathrm{h.c.}}
\end{array}
\right)  \\
-\left(
\begin{array}{c}
\mathbf{\Gamma }_{i}^{\dagger }\left( t^{\prime }\right) \mathbf{\tilde{a}}%
_{i}\left( t^{\prime }\right) \mathbf{\tilde{\rho}}\left( 0\right) \mathbf{%
\Gamma }_{j}^{\dagger }\left( t^{\prime \prime }\right) \mathbf{\tilde{a}}%
_{j}\left( t^{\prime \prime }\right) +{\mathrm{h.c.}} \\
+\mathbf{\Gamma }_{i}\left( t^{\prime }\right) \mathbf{\tilde{a}}_{i}^{%
\mathbf{\dagger }}\left( t^{\prime }\right) \mathbf{\tilde{\rho}}\left(
0\right) \mathbf{\Gamma }_{j}^{\dagger }\left( t^{\prime \prime }\right)
\mathbf{\tilde{a}}_{j}\left( t^{\prime \prime }\right) +{\mathrm{h.c.}}
\end{array}
\right)  \\
-\left(
\begin{array}{c}
\mathbf{\Gamma }_{i}^{\dagger }\left( t^{\prime }\right) \mathbf{\tilde{a}}%
_{i}\left( t^{\prime }\right) \mathbf{\tilde{\rho}}\left( 0\right) \mathbf{%
\Gamma }_{j}\left( t^{\prime \prime }\right) \mathbf{\tilde{a}}_{j}^{\mathbf{%
\dagger }}\left( t^{\prime \prime }\right) +{\mathrm{h.c.}} \\
+\mathbf{\Gamma }_{i}\left( t^{\prime }\right) \mathbf{\tilde{a}}_{i}^{%
\mathbf{\dagger }}\left( t^{\prime }\right) \mathbf{\tilde{\rho}}\left(
0\right) \mathbf{\Gamma }_{j}\left( t^{\prime \prime }\right) \mathbf{%
\tilde{a}}_{j}^{\mathbf{\dagger }}\left( t^{\prime \prime }\right)
+{\mathrm{h.c.}}
\end{array}
\right)
\end{array}
\right\} ,
\end{eqnarray}
where
\begin{equation}
\mathbf{\Gamma }_{i}\left( t\right) =\sum_{k}e^{-i\omega _{k}t}\alpha
_{ik}^{\ast }\mathbf{c}_{k},{ \ \ \ \ \ \ \ \ \ \ }\mathbf{\tilde{a}}%
_{i}\left( t\right) =e^{-i\omega _{i}t}\mathbf{a}_{i},
\end{equation}
and h.c. stands for Hermitian conjugate.

If at $t=0$ the system is prepared in state $\mathbf{\rho }_{S}\left(
0\right) $ and the environment is admitted to be in thermal equilibrium,
\begin{equation}
\mathbf{\rho }\left( 0\right) =\mathbf{\rho }_{S}\left( 0\right) \otimes
\mathbf{\rho }_{E}\left( 0\right) =\mathbf{\tilde{\rho}}\left( 0\right) ,
\end{equation}
where
\begin{eqnarray}
\mathbf{\rho }_{E}\left( 0\right) &=&\frac{1}{Z}{\prod_{k}}\exp \left(
-\beta \hbar \omega _{k}\mathbf{c}_{k}^{\dagger }\mathbf{c}_{k}\right) ={%
\prod_{k}}\left( \frac{1}{1+\bar{n}_{k}}\right) \left( \frac{\bar{n}_{k}}{1+%
\bar{n}_{k}}\right) ^{\mathbf{c}_{k}^{\dagger }\mathbf{c}_{k}},  \nonumber \\
Z &=&{\prod_{k}}\sum_{n=0}^{\infty }\exp \left( -\beta \hbar \omega
_{k}n\right) ,{ \ \ \ \ \ \ }\bar{n}_{k}=\frac{1}{\exp \left( \beta
\hbar \omega _{k}\right) -1},
\end{eqnarray}
$\beta =\left( k_{B}T\right) ^{-1}$, $k_{B}$ is Boltzmann constant, $T$ the
absolute temperature, $Z$ the partition function and $\bar{n}_{k}$ the mean
number of excitations of frequency $\omega _{k}$ \cite{Kubo1}. Taking the
limit of zero temperature, it is easy to see that
\begin{eqnarray}
\mathbf{\tilde{\rho}}_{S}\left( t\right) -\mathbf{\tilde{\rho}}_{S}\left(
0\right) &=&\sum_{i,j=1}^{2}{\int_{0}^{t}}dt^{\prime }{\int_{0}^{\ t\prime }}%
d\tau \left\{ \left( \sum_{k}\alpha _{ik}\alpha _{jk}^{\ast }e^{i\omega
_{k}\tau }\right) e^{-i\omega _{j}\tau }e^{i\left( \omega _{j}-\omega
_{i}\right) t^{\prime }}\left( \mathbf{a}_{i}\mathbf{\tilde{\rho}}_{S}\left(
0\right) \mathbf{a}_{j}^{\mathbf{\dagger }}-\mathbf{\tilde{\rho}}_{S}\left(
0\right) \mathbf{a}_{j}^{\mathbf{\dagger }}\mathbf{a}_{i}\right) \right\}
\nonumber \\
&&+{\mathrm{h.c.}},
\end{eqnarray}
We used Fock states basis to take the traces over the environment and used
for the integrals $\tau =t^{\prime }-t^{\prime \prime }$.

Notice that $\sum_{k}\alpha _{ik}\alpha _{jk}^{\ast }e^{i\omega _{k}\tau }$
decays very fast when $\tau $ grows. Considering a huge number of
environment modes, with frequencies spread over a large interval of values
and a weak coupling to each mode (small $\alpha _{ik}$), the phases of the $%
e^{i\omega _{k}\tau }$ make the sums practically null when $\tau $ is not
close to zero. If $\tau _{c}$ is the time when $\sum_{k}\alpha _{ik}\alpha
_{jk}^{\ast }e^{i\omega _{k}\tau }$ have appreciable values, we can, for $%
t\gg \tau _{c}$, modify the integration limits without changing
substantially the result. Thus we obtain
\begin{equation}
\mathbf{\tilde{\rho}}_{S}\left( t\right) -\mathbf{\tilde{\rho}}_{S}\left(
0\right) =\sum_{i,j=1}^{2}\left\{
\begin{array}{c}
\left( k_{ij}+i\Delta _{ij}\right) \left( \mathbf{a}_{i}\mathbf{\tilde{\rho}}%
_{S}\left( 0\right) \mathbf{a}_{j}^{\mathbf{\dagger }}-\mathbf{\tilde{\rho}}%
_{S}\left( 0\right) \mathbf{a}_{j}^{\mathbf{\dagger }}\mathbf{a}_{i}\right)
\left( {\int_{0}^{t}}dt^{\prime }e^{i\left( \omega _{j}-\omega _{i}\right)
t^{\prime }}\right)
\end{array}
\right\} +{\mathrm{h.c.}},  \label{ros1}
\end{equation}
where the real numbers $k_{ij}$ and $\Delta _{ij}$ are given by
\begin{equation}
k_{ij}+i\Delta _{ij}=\sum_{k}\alpha _{ik}\alpha _{jk}^{\ast }{\int_{0}^{\tau
_{c}}}d\tau e^{i\left( \omega _{k}-\omega _{j}\right) \tau }.
\label{k_ij+delta_ij}
\end{equation}

Differentiating both sides of eq. (\ref{ros1}) and iterating leads to
\begin{equation}
\frac{d}{dt}\mathbf{\tilde{\rho}}_{S}\left( t\right)
=\sum_{i,j=1}^{2}\left\{
\begin{array}{c}
\left( k_{ij}+i\Delta _{ij}\right) \left( \mathbf{a}_{i}\mathbf{\tilde{\rho}}%
_{S}\left( t\right) \mathbf{a}_{j}^{\mathbf{\dagger }}-\mathbf{\tilde{\rho}}%
_{S}\left( t\right) \mathbf{a}_{j}^{\mathbf{\dagger }}\mathbf{a}_{i}\right)
e^{i\left( \omega _{j}-\omega _{i}\right) t} \\
+\mathcal{O}^{2}(k_{ij}+i\Delta _{ij})
\end{array}
\right\} +{\mathrm{h.c.}}.
\end{equation}
The terms $\mathcal{O}^{2}(k_{ij}+i\Delta _{ij})$ are similar to the ones
that appear if we consider $\mathcal{O}^{4}\left( \tilde{H}_{int}\right) $
terms. Since we disregard $\mathcal{O}^{3}\left( \tilde{H}_{int}\right) $,
we will not take them in to account.

With the help of
\begin{equation}
\frac{d}{dt}\mathbf{\tilde{\rho}}_{S}\left( t\right) =e^{\frac{i}{\hbar }%
\mathbf{H}_{S}t}\left\{ \frac{i}{\hbar }\left[ \mathbf{H}_{S},\mathbf{\rho }%
_{S}\left( t\right) \right] +\frac{d}{dt}\mathbf{\rho }_{S}\left( t\right)
\right\} e^{-\frac{i}{\hbar }\mathbf{H}_{S}t}
\end{equation}
we return to Schr\"{o}dinger picture and write the master equation
\begin{equation}
\frac{d}{dt}\mathbf{\rho }_{S}\left( t\right) =\mathcal{L}\mathbf{\rho }%
_{S}\left( t\right) ,
\end{equation}
where the Liouvillian superoperator (\textit{i.e.}: operator which acts on
operators) can be decomposed in three parts,
\begin{equation}
\mathcal{L=L}_{1}+\mathcal{L}_{2}+\mathcal{L}_{12},
\end{equation}
and the parts are given by
\begin{eqnarray}
\mathcal{L}_{1} &=&k_{11}\left( 2\mathbf{a}_{1}\bullet \mathbf{a}_{1}^{%
\mathbf{\dagger }}-\bullet \mathbf{a}_{1}^{\mathbf{\dagger }}\mathbf{a}_{1}-%
\mathbf{a}_{1}^{\mathbf{\dagger }}\mathbf{a}_{1}\bullet \right) +i\left(
\Delta _{11}-\omega _{1}\right) \left[ \mathbf{a}_{1}^{\mathbf{\dagger }}%
\mathbf{a}_{1},\bullet \right]  \nonumber \\
\mathcal{L}_{2} &=&k_{22}\left( 2\mathbf{a}_{2}\bullet \mathbf{a}_{2}^{%
\mathbf{\dagger }}-\bullet \mathbf{a}_{2}^{\mathbf{\dagger }}\mathbf{a}_{2}-%
\mathbf{a}_{2}^{\mathbf{\dagger }}\mathbf{a}_{2}\bullet \right) +i\left(
\Delta _{22}-\omega _{2}\right) \left[ \mathbf{a}_{2}^{\mathbf{\dagger }}%
\mathbf{a}_{2},\bullet \right]  \nonumber \\
\mathcal{L}_{12} &=&k_{12}\left( \mathbf{a}_{1}\bullet \mathbf{a}_{2}^{%
\mathbf{\dagger }}+\mathbf{a}_{2}\bullet \mathbf{a}_{1}^{\mathbf{\dagger }%
}-\bullet \mathbf{a}_{2}^{\mathbf{\dagger }}\mathbf{a}_{1}-\mathbf{a}_{1}^{%
\mathbf{\dagger }}\mathbf{a}_{2}\bullet \right) +  \nonumber \\
&&k_{21}\left( \mathbf{a}_{2}\bullet \mathbf{a}_{1}^{\mathbf{\dagger }}+%
\mathbf{a}_{1}\bullet \mathbf{a}_{2}^{\mathbf{\dagger }}-\bullet \mathbf{a}%
_{1}^{\mathbf{\dagger }}\mathbf{a}_{2}-\mathbf{a}_{2}^{\mathbf{\dagger }}%
\mathbf{a}_{1}\bullet \right) +  \nonumber \\
&&i\left( \frac{\Delta _{12}-\Delta _{21}}{2}\right) \left( \mathbf{a}%
_{1}\bullet \mathbf{a}_{2}^{\mathbf{\dagger }}-\mathbf{a}_{2}\bullet \mathbf{%
a}_{1}^{\mathbf{\dagger }}-\bullet \mathbf{a}_{2}^{\mathbf{\dagger }}\mathbf{%
a}_{1}+\mathbf{a}_{1}^{\mathbf{\dagger }}\mathbf{a}_{2}\bullet \right) +
\nonumber \\
&&i\left( \frac{\Delta _{21}-\Delta _{12}}{2}\right) \left( \mathbf{a}%
_{2}\bullet \mathbf{a}_{1}^{\mathbf{\dagger }}-\mathbf{a}_{1}\bullet \mathbf{%
a}_{2}^{\mathbf{\dagger }}-\bullet \mathbf{a}_{1}^{\mathbf{\dagger }}\mathbf{%
a}_{2}+\mathbf{a}_{2}^{\mathbf{\dagger }}\mathbf{a}_{1}\bullet \right) +
\nonumber \\
&&i\left( \frac{\Delta _{12}+\Delta _{21}}{2}\right) \left[ \mathbf{a}_{1}^{%
\mathbf{\dagger }}\mathbf{a}_{2}+\mathbf{a}_{2}^{\mathbf{\dagger }}\mathbf{a}%
_{1},\bullet \right] .
\end{eqnarray}
A word about notation is in order: we use the conventional notation for
superoperators \cite{Faria1} where the dot ($\bullet $) indicates the place
to be occupied by the operator on which the superoperator is acting.

Now we consider the environment large enough to take the limit of dense
spectrum. In this sense we define a density of modes $D\left( \omega \right)
$ for the environment, where $D\left( \omega \right) d\omega $ represents
the number of modes in the frequency interval between $\omega $ and $\omega
+d\omega $. This allows the conversion of sums over modes $\left(
\sum_{k}\right) $ into integrals over frequencies $\left( \int D\left(
\omega \right) d\omega \right) $, with the corresponding change of $\alpha
_{k}$ for $\alpha \left( \omega \right) $ and $\beta _{k}$ for $\beta \left(
\omega \right) $, leading to
\begin{eqnarray}
k_{ij}+i\Delta _{ij} &=&\int d\omega D\left( \omega \right) \alpha
_{i}\left( \omega \right) \alpha _{j}^{\ast }\left( \omega \right) \xi _{j},
\nonumber \\
\xi _{j} &=&{\int_{0}^{\tau _{c}}}d\tau e^{i\left( \omega -\omega
_{j}\right) \tau }.  \label{coefficientsnm}
\end{eqnarray}

Using eqs. (\ref{transf}) we may write $\mathcal{L}$ in terms of the
operators related to the original laboratory modes $A$ and $B$:
\begin{eqnarray}
\mathcal{L} &=&\mathcal{L}_{A}+\mathcal{L}_{B}+\mathcal{L}_{int},  \nonumber
\\
\mathcal{L}_{A} &=&k_{aa}\left( 2\mathbf{a}\bullet \mathbf{a}^{\mathbf{%
\dagger }}-\bullet \mathbf{a}^{\mathbf{\dagger }}\mathbf{a}-\mathbf{a}^{%
\mathbf{\dagger }}\mathbf{a}\bullet \right) +i\left( \Delta _{aa}-\omega
_{a}\right) \left[ \mathbf{a}^{\mathbf{\dagger }}\mathbf{a},\bullet \right]
\nonumber \\
\mathcal{L}_{B} &=&k_{bb}\left( 2\mathbf{b}\bullet \mathbf{b}^{\mathbf{%
\dagger }}-\bullet \mathbf{b}^{\mathbf{\dagger }}\mathbf{b}-\mathbf{b}^{%
\mathbf{\dagger }}\mathbf{b}\bullet \right) +i\left( \Delta _{bb}-\omega
_{b}\right) \left[ \mathbf{b}^{\mathbf{\dagger }}\mathbf{b},\bullet \right]
\nonumber \\
\mathcal{L}_{int} &=&k_{ab}\left( \mathbf{a}\bullet \mathbf{b}^{\mathbf{%
\dagger }}+\mathbf{b}\bullet \mathbf{a}^{\mathbf{\dagger }}-\bullet \mathbf{b%
}^{\mathbf{\dagger }}\mathbf{a-a}^{\mathbf{\dagger }}\mathbf{b}\bullet
\right) +  \nonumber \\
&&k_{ba}\left( \mathbf{b}\bullet \mathbf{a}^{\mathbf{\dagger }}+\mathbf{a}%
\bullet \mathbf{b}^{\mathbf{\dagger }}-\bullet \mathbf{a}^{\mathbf{\dagger }}%
\mathbf{b-b}^{\mathbf{\dagger }}\mathbf{a}\bullet \right) +  \nonumber \\
&&i\left( \frac{\Delta _{ab}-\Delta _{ba}}{2}\right) \left( \mathbf{a}%
\bullet \mathbf{b}^{\mathbf{\dagger }}-\mathbf{b}\bullet \mathbf{a}^{\mathbf{%
\dagger }}-\bullet \mathbf{b}^{\mathbf{\dagger }}\mathbf{a+a}^{\mathbf{%
\dagger }}\mathbf{b}\bullet \right) +  \nonumber \\
&&i\left( \frac{\Delta _{ba}-\Delta _{ab}}{2}\right) \left( \mathbf{b}%
\bullet \mathbf{a}^{\mathbf{\dagger }}-\mathbf{a}\bullet \mathbf{b}^{\mathbf{%
\dagger }}-\bullet \mathbf{a}^{\mathbf{\dagger }}\mathbf{b+b}^{\mathbf{%
\dagger }}\mathbf{a}\bullet \right) +  \nonumber \\
&&i\left( \left( \frac{\Delta _{ab}+\Delta _{ba}}{2}\right) -g\right) \left[
\mathbf{b}^{\mathbf{\dagger }}\mathbf{a+\mathbf{a}^{\mathbf{\dagger }}%
\mathbf{b},}\bullet \right] .  \label{master}
\end{eqnarray}
Here the coefficients are
\begin{eqnarray}
k_{aa}+i\Delta _{aa} &=&\int d\omega D\left( \omega \right) \left\{ \alpha
\left( \omega \right) \alpha ^{\ast }\left( \omega \right) \eta +\alpha
\left( \omega \right) \beta ^{\ast }\left( \omega \right) \nu \right\} ,
\label{k_aa+idelta_aa} \\
k_{ab}+i\Delta _{ab} &=&\int d\omega D\left( \omega \right) \left\{ \alpha
\left( \omega \right) \alpha ^{\ast }\left( \omega \right) \nu +\alpha
\left( \omega \right) \beta ^{\ast }\left( \omega \right) \mu \right\} ,
\label{k_ab+idelta_ab} \\
k_{ba}+i\Delta _{ba} &=&\int d\omega D\left( \omega \right) \left\{ \beta
\left( \omega \right) \alpha ^{\ast }\left( \omega \right) \eta +\beta
\left( \omega \right) \beta ^{\ast }\left( \omega \right) \nu \right\} ,
\label{k_ba+idelta_ba} \\
k_{bb}+i\Delta _{bb} &=&\int d\omega D\left( \omega \right) \left\{ \beta
\left( \omega \right) \alpha ^{\ast }\left( \omega \right) \nu +\beta \left(
\omega \right) \beta ^{\ast }\left( \omega \right) \mu \right\} ,
\label{k_bb+idelta_bb} \\
\eta &=&\xi _{1}\cos ^{2}\theta +\xi _{2}\sin ^{2}\theta ,{ \ \ \ \ }%
\mu =\xi _{2}\cos ^{2}\theta +\xi _{1}\sin ^{2}\theta ,{ \ \ \ \ }\nu
=\left( \xi _{1}-\xi _{2}\right) \sin \theta \cos \theta .  \label{auxcoe}
\end{eqnarray}
It is important to note that all superoperators appearing in eqs. (\ref
{master}) form a closed Lie algebra (see table 1).

$\mathcal{L}_{A}$ and $\mathcal{L}_{B}$ are formally similar to the
Liouvillians of independent modes evolving subjected to environment and $%
\mathcal{L}_{int}$ has the form of an ``interaction Liouvillian''. However
the constants $k_{aa}$ and $\Delta _{aa}$ appearing in $\mathcal{L}_{A}$
depend on the coupling constants to the environment modes of both modes, $%
\alpha \left( \omega \right) $ and $\beta \left( \omega \right) $ (and
analogously for $k_{bb}$ and $\Delta _{bb}$). The $\alpha \left( \omega
\right) \beta ^{\ast }\left( \omega \right) $ and $\beta \left( \omega
\right) \alpha ^{\ast }\left( \omega \right) $ terms appearing in eqs. (\ref
{k_aa+idelta_aa}) and (\ref{k_bb+idelta_bb}) are related to the return from
the normal modes when we write $\mathcal{L}$ in terms of operators of the
original laboratory modes. Notice that (eqs. (\ref{coefficientsnm})) $k_{ii}$
and $\Delta _{ii}$ ($i=1$ or $2$) depend only on $\left| \alpha _{i}\left(
\omega \right) \right| ^{2}$. In fact, it is more natural to see the whole
system through its normal modes, and not to look at each cavity mode. This
is what the environment does: due to the form of $\xi _{1}$ and $\xi _{2}$,
the values of $\alpha \left( \omega \right) $, $\beta \left( \omega \right) $
and $D\left( \omega \right) $ that effectively contribute to the
coefficients of eqs. (\ref{master}) are the ones with frequencies around the
normal modes frequencies $\omega _{1}$ and $\omega _{2}$, and not around the
original laboratory modes frequencies $\omega _{a}$ and $\omega _{b}$.

The physical interpretation for the coefficients $\Delta _{xy}$ and $k_{xy}$
($x$, $y=a$ or $b$) is straightforward if $x=y$: $\Delta _{aa}$ and $\Delta
_{bb}$ concern the system's unitary evolution, changing the oscillation
frequencies of the cavities (the Lamb shift); $k_{aa}$ and $k_{bb}$ are
dissipation constants. For $x\neq y$ the coefficients are related to a
communication channel between the cavities mediated by the environment,
playing unitary and non unitary roles in the dynamics (notice that $%
k_{ab}+i\Delta _{ab}$ and $k_{ba}+i\Delta _{ba}$ are not necessarily equal
nor complex conjugate). As will become clear in sections (3) and (5), the
values of $k_{ab}$ and $k_{ba}$ are very important for decoherence control.
In eqs. (\ref{k_ab+idelta_ab}) and (\ref{k_ba+idelta_ba}) we find terms with
$\alpha \left( \omega \right) \beta ^{\ast }\left( \omega \right) $ and $%
\beta \left( \omega \right) \alpha ^{\ast }\left( \omega \right) $ which
integrals tend to be null if the phases of $\alpha \left( \omega \right) $ and 
$\beta \left( \omega \right) $ are not correlated. It is probably the most
common case for fields in cavities. We must emphasize that analogous situation
were studied in refs. \cite{Akram1} and \cite{Ficek1} where the authors discuss
the problem of two electric dipoles interacting with the vacuum of
electromagnetic field. The situation is formally similar, just changing
bosonic operators for spin operators. However, as the dipoles are the only
possibility of coupling, that system naturally exhibit this phase relation,
and one can show \cite{Romero1,Kim1} that this system is equivalent to the
dipoles only coupled to one dissipative field mode. In the case of cavities,
although it is desirable the phase relation, even in the uncorrelated case $%
k_{ab}$ and $k_{ba}$ may have appreciable values if $\nu \neq 0$ (eq. (\ref
{auxcoe})), which may be achievable if, for example, the density of modes
around $\omega _{1}$ have a different behavior than around $\omega _{2}$. It
is another effect that appear when we write eq. (\ref{master}), going back
to the original laboratory modes. Since the environment here is composed
mainly by phonons in the mirrors of the cavities and by electromagnetic
modes in the space around the cavities, it is not impossible to manipulate
the density of modes.

\section{General solution}

\label{Gensol}Before going over to specific examples, we present in detail
the general solution of the master equation (\ref{master}).

The master equation (\ref{master}) can be solved using Witschel's technique
\cite{Witschel1} and the parameter derivation technique\cite{Wilcox1}. This
technique allows one to determine coefficients $\delta _{i}$ such that the
following identity is valid
\begin{equation}
e^{\left( \gamma _{1}\mathbf{A}_{1}+\gamma _{2}\mathbf{A}_{2}+\cdots +\gamma
_{n}\mathbf{A}_{n}\right) t}=e^{\delta _{1}\left( t\right) \mathbf{A}%
_{1}}e^{\delta _{2}\left( t\right) \mathbf{A}_{2}}\cdots e^{\delta
_{n}\left( t\right) \mathbf{A}_{n}}  \label{exp(L0t)=exp(d1A1)...exp(dnAn)}
\end{equation}
where the $\mathbf{A}_{i}$'s are superoperators forming a closed Lie algebra
and $t$ is a parameter (in our case, time). The parameter derivation
technique consists of the following procedure:

\begin{enumerate}
\item  {Derive both sides of eq. (\ref{exp(L0t)=exp(d1A1)...exp(dnAn)}) with
respect to $t$ and get
\begin{eqnarray}
\left( \sum_{i=1}^{n}\gamma _{i}\mathbf{A}_{i}\right) \exp \left(
\sum_{i=1}^{n}\gamma _{i}\mathbf{A}_{i}t\right) &=&\dot{\delta}_{1}\left(
t\right) \mathbf{A}_{1}\prod_{i=1}^{n}e^{\delta _{i}\left( t\right) \mathbf{A%
}_{i}}  \nonumber \\
&&+\dot{\delta}_{2}\left( t\right) e^{\delta _{1}\left( t\right) \mathbf{A}%
_{1}}\mathbf{A}_{2}\prod_{i=2}^{n}e^{\delta _{i}\left( t\right) \mathbf{A}%
_{i}}  \nonumber \\
&&+\cdots  \nonumber \\
&&+\dot{\delta}_{n}\left( t\right) \prod_{i=1}^{n-1}e^{\delta _{i}\left(
t\right) \mathbf{A}_{i}}\mathbf{A}_{n}e^{\delta _{n}\left( t\right) \mathbf{A%
}_{n}}.  \label{(d/dt)exp(L0t)}
\end{eqnarray}
}

\item  {\label{step2}Use the similarity transformation
\begin{equation}
e^{x\mathbf{B}}\mathbf{A}e^{-x\mathbf{B}}=e^{x\left[ \mathbf{B},\bullet %
\right] }\mathbf{A},
\end{equation}
and the linear independence of $\left\{ \mathbf{A}_{i}\right\} $ in order to
obtain differential equations for the parameters $\delta _{i}\left( t\right)
$.}

\item  {Solve the $c$-numbers differential equations and obtain the
factorized evolution superoperator written in the right hand side\ of eq. (%
\ref{exp(L0t)=exp(d1A1)...exp(dnAn)}).}
\end{enumerate}

As an example of step \ref{step2} above, let us take the second term in the
r.h.s. of eq. (\ref{(d/dt)exp(L0t)})
\begin{eqnarray}
e^{\delta _{1}\left( t\right) \mathbf{A}_{1}}\mathbf{A}_{2} &=&e^{\delta
_{1}\left( t\right) \mathbf{A}_{1}}\mathbf{A}_{2}e^{-\delta _{1}\left(
t\right) \mathbf{A}_{1}}e^{\delta _{1}\left( t\right) \mathbf{A}_{1}}
\nonumber \\
&=&e^{\delta _{1}\left( t\right) \left[ \mathbf{A}_{1},\bullet \right] }%
\mathbf{A}_{2}e^{\delta _{1}\left( t\right) \mathbf{A}_{1}}  \nonumber \\
&\equiv &F_{2}\left( \delta _{1}\left( t\right) ,\{\mathbf{A}_{i}\}\right)
e^{\delta _{1}\left( t\right) \mathbf{A}_{1}}.
\end{eqnarray}
Analogously one can define a similar operation for all the other terms and
define
\begin{equation}
F_{3}\equiv F_{3}\left( \delta _{1}\left( t\right) ,\delta _{2}\left(
t\right) ,\{\mathbf{A}_{i}\}\right) ,\cdots ,F_{n}\equiv F_{3}\left( \delta
_{1}\left( t\right) ,\delta _{2}\left( t\right) ,\cdots ,\delta _{n-1}\left(
t\right) ,\{\mathbf{A}_{i}\}\right) ,
\end{equation}
and write
\begin{equation}
\left( \sum_{i=1}^{n}\gamma _{i}\mathbf{A}_{i}\right) \exp \left(
\sum_{i=1}^{n}\gamma _{i}\mathbf{A}_{i}t\right) =\left( \dot{\delta}_{1}%
\mathbf{A}_{1}+\dot{\delta}_{2}F_{2}+\cdots +\dot{\delta}_{n}F_{n}\right)
\exp \left( \sum_{i=1}^{n}\gamma _{i}\mathbf{A}_{i}t\right)
\end{equation}
or, equivalently,
\begin{equation}
\sum_{i=1}^{n}\gamma _{i}\mathbf{A}_{i}=\dot{\delta}_{1}\mathbf{A}_{1}+\dot{%
\delta}_{2}F_{2}+\cdots +\dot{\delta}_{n}F_{n}.
\end{equation}
Comparing coefficients of each $\mathbf{A}_{i}$ (due to linear independence)
one then obtains a system of coupled differential equations for the $\delta
_{i}$'s.

In our case we have
\begin{eqnarray}
\mathbf{\rho }_{S}\left( t\right) &=&e^{\mathcal{L}t}\mathbf{\rho }%
_{S}\left( 0\right)  \nonumber \\
&=&e^{h_{1}\left( t\right) \mathbf{a\bullet a}^{\dagger }}e^{h_{2}\left(
t\right) \mathbf{b\bullet b}^{\dagger }}e^{z_{l}\left( t\right) \mathbf{%
a\bullet b}^{\dagger }}e^{z\left( t\right) \mathbf{b\bullet a}^{\dagger
}}e^{n_{l}\left( t\right) \mathbf{\bullet a}^{\dagger }\mathbf{b}}e^{n\left(
t\right) \mathbf{b}^{\dagger }\mathbf{a\bullet }}  \nonumber \\
&&e^{m_{2}\left( t\right) \mathbf{b}^{\dagger }\mathbf{b\bullet }%
}e^{p_{2}\left( t\right) \mathbf{\bullet b}^{\dagger }\mathbf{b}%
}e^{m_{1}\left( t\right) \mathbf{a}^{\dagger }\mathbf{a\bullet }%
}e^{p_{1}\left( t\right) \mathbf{\bullet a}^{\dagger }\mathbf{a}}e^{q\left(
t\right) \mathbf{a}^{\dagger }\mathbf{b\bullet }}e^{q_{l}\left( t\right)
\mathbf{\bullet b}^{\dagger }\mathbf{a}}\mathbf{\rho }_{S}\left( 0\right) .
\end{eqnarray}

Using the method just described we get
\begin{eqnarray}
-i\Omega _{aa}-k_{aa} &=&\dot{m}_{1}\left( t\right) -n\left( t\right) \dot{q}%
\left( t\right) e^{m_{1}\left( t\right) -m_{2}\left( t\right) },  \nonumber
\\
-i\Omega _{bb}-k_{bb} &=&\dot{m}_{2}\left( t\right) +n\left( t\right) \dot{q}%
\left( t\right) e^{m_{1}\left( t\right) -m_{2}\left( t\right) },  \nonumber
\\
-i\Omega _{ab}-k_{ab} &=&\dot{q}\left( t\right) e^{m_{1}\left( t\right)
-m_{2}\left( t\right) },  \nonumber \\
-i\Omega _{ba}-k_{ba} &=&\dot{n}\left( t\right) +n\left( t\right) \left(
\dot{m}_{1}\left( t\right) -\dot{m}_{2}\left( t\right) \right) -n\left(
t\right) ^{2}\dot{q}\left( t\right) e^{m_{1}\left( t\right) -m_{2}\left(
t\right) },  \nonumber \\
+i\Omega _{aa}-k_{aa} &=&\dot{p}_{1}\left( t\right) -n_{l}\left( t\right)
\dot{q}_{l}\left( t\right) e^{p_{1}\left( t\right) -p_{2}\left( t\right) },
\nonumber \\
+i\Omega _{bb}-k_{bb} &=&\dot{p}_{2}\left( t\right) +n_{l}\left( t\right)
\dot{q}_{l}\left( t\right) e^{p_{1}\left( t\right) -p_{2}\left( t\right) },
\nonumber \\
+i\Omega _{ab}-k_{ab} &=&\dot{q}_{l}\left( t\right) e^{p_{1}\left( t\right)
-p_{2}\left( t\right) },  \nonumber \\
+i\Omega _{ba}-k_{ba} &=&\dot{n}_{l}\left( t\right) +n_{l}\left( t\right)
\left( \dot{p}_{1}\left( t\right) -\dot{p}_{2}\left( t\right) \right)
-n_{l}\left( t\right) ^{2}\dot{q}_{l}\left( t\right) e^{p_{1}\left( t\right)
-p_{2}\left( t\right) },  \nonumber \\
2k_{aa} &=&z\left( t\right) \left( -i\Omega _{ba}-k_{ba}\right)  \nonumber \\
&&+z_{l}\left( t\right) \left( +i\Omega _{ba}-k_{ba}\right) +h_{1}\left(
t\right) \left( -2k_{aa}\right) +\dot{h_{1}}\left( t\right) ,  \nonumber \\
2k_{bb} &=&z\left( t\right) \left( +i\Omega _{ab}-k_{ab}\right)  \nonumber \\
&&+z_{l}\left( t\right) \left( -i\Omega _{ab}-k_{ab}\right) +h_{2}\left(
t\right) \left( -2k_{bb}\right) +\dot{h_{2}}\left( t\right) ,  \nonumber \\
+i\left( \Omega _{ab}-\Omega _{ba}\right) +k_{ba}+k_{ab} &=&z\left( t\right)
\left( +i\Omega _{aa}-i\Omega _{bb}-k_{aa}-k_{bb}\right)  \nonumber \\
&&+h_{1}\left( t\right) \left( -i\Omega _{ab}-k_{ab}\right) +h_{2}\left(
t\right) \left( +i\Omega _{ba}-k_{ba}\right) +\dot{z}\left( t\right) ,
\nonumber \\
+i\left( \Omega _{ba}-\Omega _{ab}\right) +k_{ab}+k_{ba} &=&z_{l}\left(
t\right) \left( -i\Omega _{aa}+i\Omega _{bb}-k_{aa}-k_{bb}\right)  \nonumber
\\
&&+h_{1}\left( t\right) \left( +i\Omega _{ab}-k_{ab}\right) +h_{2}\left(
t\right) \left( -i\Omega _{ba}-k_{ba}\right) +\dot{z_{l}}\left( t\right) .
\end{eqnarray}
where
\begin{equation}
\Omega _{aa}=\omega _{a}-\Delta _{aa},{ \ \ \ \ }\Omega _{bb}=\omega
_{b}-\Delta _{bb},{ \ \ \ \ }\Omega _{ab}=g-\Delta _{ab},{ \ \ \ \
}\Omega _{ba}=g-\Delta _{ba}.
\end{equation}
The solution reads
\begin{eqnarray}
n\left( t\right) &=&\frac{l_{2}\left( t\right) }{f_{1}\left( t\right) }%
,\quad \quad q\left( t\right) =\frac{l_{1}\left( t\right) }{f_{1}\left(
t\right) },  \nonumber \\
e^{m_{1}\left( t\right) } &=&e^{-Rt}f_{1}\left( t\right) ,\quad \quad
e^{m_{2}\left( t\right) }=e^{-2Rt}e^{-m_{1}\left( t\right) },  \nonumber \\
h_{1}\left( t\right) &=&\frac{\left| f_{2}\left( t\right) \right|
^{2}+\left| l_{2}\left( t\right) \right| ^{2}}{e^{-2k_{m}t}}-1,  \nonumber \\
h_{2}\left( t\right) &=&\frac{\left| f_{1}\left( t\right) \right|
^{2}+\left| l_{1}\left( t\right) \right| ^{2}}{e^{-2k_{m}t}}-1,  \nonumber \\
z\left( t\right) &=&\frac{-l_{1}\left( t\right) f_{2}^{\ast }\left( t\right)
-l_{2}^{\ast }\left( t\right) f_{1}\left( t\right) }{e^{-2k_{m}t}},
\nonumber \\
z_{l}\left( t\right) &=&z^{\ast }\left( t\right) ,  \nonumber \\
n_{l}\left( t\right) &=&\left( n\left( t\right) \right) ^{\ast },\quad \quad
q_{l}\left( t\right) =\left( q\left( t\right) \right) ^{\ast },  \nonumber \\
p_{2}\left( t\right) &=&\left( m_{2}\left( t\right) \right) ^{\ast },\quad
\quad p_{1}\left( t\right) =\left( m_{1}\left( t\right) \right) ^{\ast },
\end{eqnarray}
where
\begin{eqnarray}
f_{1}\left( t\right) &=&\frac{1}{2r}\left[ \left( r+c\right) e^{rt}+\left(
r-c\right) e^{-rt}\right] ,  \nonumber \\
f_{2}\left( t\right) &=&\frac{1}{2r}\left[ \left( r-c\right) e^{rt}+\left(
r+c\right) e^{-rt}\right] ,  \nonumber \\
l_{1}\left( t\right) &=&\frac{1}{2r}\left( i\Omega _{ab}+k_{ab}\right)
\left( e^{-rt}-e^{rt}\right) ,  \nonumber \\
l_{2}\left( t\right) &=&\frac{1}{2r}\left( i\Omega _{ba}+k_{ba}\right)
\left( e^{-rt}-e^{rt}\right) ,
\end{eqnarray}
and
\begin{eqnarray}
c &=&\frac{k_{bb}-k_{aa}}{2}+i\frac{\left( \Omega _{bb}-\Omega _{aa}\right)
}{2},  \nonumber \\
r &=&\sqrt{c^{2}+\left( i\Omega _{ba}+k_{ba}\right) \left( i\Omega
_{ab}+k_{ab}\right) },  \nonumber \\
R &=&\frac{k_{bb}+k_{aa}}{2}+i\frac{\Omega _{aa}+\Omega _{bb}}{2}%
=k_{m}+i\omega _{m}.  \label{const}
\end{eqnarray}

The above equations look complicated. However there are some ingredients in
them which become very illuminating from the physical point of view. Let us
then discuss a few of them in a general context and apply the results in
specific situations in the next two sections.

It is probably sufficient to discuss the equation
\begin{equation}
e^{m_{1}\left( t\right) }=\frac{1}{2r}e^{-Rt}\left[ \left( r+c\right)
e^{rt}+\left( r-c\right) e^{-rt}\right] ,  \label{todisc}
\end{equation}
since it carries the essential ingredients of dynamical evolution. The point
we want to stress is that by controlling some parameters, one can have
distinct regimes of dissipation and decoherence in this model. The factor $%
e^{-Rt}$ in eq.\ (\ref{todisc}) gives the expected decay behavior, with the
average dissipation rate $k_{m}$. The term $e^{-rt}$ contains two
contributions: a term which measures to which degree the two cavities are
different, $c$, which adds to the other term coming from the dynamical
coupling of the modes through the environment ($k_{ab}-i\Delta _{ab}$ and $%
k_{ba}-i\Delta _{ba}$) and directly, \textit{i.e.}: ``through an antenna'', (%
$ig$). One interesting effect is that for some parameter choices, this $%
e^{rt}$ term can cancel the deleterious effects of the overall factor $%
e^{-Rt}$. For example, take $\Omega _{aa}=\Omega _{bb}$ and $g=0$, \textit{%
i.e.}: two degenerated modes. Note that if $\left( k_{ab}-i\Delta
_{ab}\right) \left( k_{ba}-i\Delta _{ba}\right) =k_{aa}k_{bb}$, we get $%
r=k_{m}$, and, in this term, dissipation can be partially suppressed.
Decoherence control is a little more intrincated matter, since it also
depends on initial states as well. This will be discussed shortly.

\section{Two weakly coupled cavities with initial coherent states:
manipulating dissipation}

\label{Cohstates} In order to analyze specific cases, let us start with the
simple situation in which coherent states are initially fed into the
cavities:
\begin{equation}
\mathbf{\rho }_{S}\left( 0\right) =\left| v_{a}\right\rangle \left\langle
v_{a}\right| \otimes \left| v_{b}\right\rangle \left\langle v_{b}\right| .
\end{equation}
Applying the results of the previous section we get for the time evolution
of this state
\begin{equation}
\mathbf{\rho }_{S}\left( t\right) =\left| v_{a}\left( t\right) \right\rangle
\left\langle v_{a}\left( t\right) \right| \otimes \left| v_{b}\left(
t\right) \right\rangle \left\langle v_{b}\left( t\right) \right| ,
\label{cohert}
\end{equation}
where
\begin{eqnarray}
v_{a}\left( t\right) &=&e^{-\left( k_{m}+i\omega _{m}\right) t}\left\{
v_{a}f_{1}\left( t\right) +v_{b}l_{1}\left( t\right) \right\} ,  \nonumber \\
v_{b}\left( t\right) &=&e^{-\left( k_{m}+i\omega _{m}\right) t}\left\{
v_{a}l_{2}\left( t\right) +v_{b}f_{2}\left( t\right) \right\} .
\end{eqnarray}
The first point to stress is that the known result of dissipative zero
temperature time evolution of coherent states linearly coupled to the
environment gives coherent states \cite{Zurek1} applies here. Another point
is that the auxiliary functions $f_{i}$ and $l_{i}$ are oscillatory,
generalizing the well known behavior of cosine and sine in simple cases like
two identical isolated modes. Finally, as it should be, the symmetric
character of the expressions: each mode fed the other with a strength
depending on the coupling $i\Omega _{ab}+k_{ab}$ and $i\Omega _{ba}+k_{ba}$.

In order to gain further insight let us make some simplifying but realistic
assumptions. Consider two modes in different cavities ($A$ and $B$) with the
same frequency, $\omega _{a}=\omega _{b}=\omega $. The cavities are distant
in the scale of the wavelength of the relevant mode, what leads to
uncorrelated $\alpha \left( \omega \right) $ and $\beta \left( \omega
\right) $ (see section (6)) and then $\int D\left( \omega \right) \alpha
\left( \omega \right) \beta ^{\ast }\left( \omega \right) \xi _{i}=0$ ($i=1$
and $2$). If we have also $D\left( \omega \right) $, $\left| \alpha \left(
\omega \right) \right| $ and $\left| \beta \left( \omega \right) \right| $
slow varying around $\omega _{1}$ and $\omega _{2}$, the lamb
shifts will be negligible: $\Delta _{aa}=\Delta _{bb}=0$. Consider now very
different dissipation rates, $k_{bb}\gg k_{aa}$ (two geometrically identical
cavities made of different materials, \textit{e.g.}: one superconductor and
the other metallic), and weak direct coupling in the sense $g\ll \Delta k$ ($%
\Delta k=k_{bb}-k_{aa}$). Small $g$ implies $\omega _{1}\simeq \omega
_{2}\simeq \omega $. Thus $\nu \simeq 0$ and, since we already have $\int
D\left( \omega \right) \alpha \left( \omega \right) \beta ^{\ast }\left(
\omega \right) \xi _{i}=0$, the coupling through the environment practically
vanishes: $k_{ab}-i\Delta _{ab}=k_{ba}-i\Delta _{ba}=0$. These conditions
lead to
\begin{equation}
\frac{c}{r}=\frac{\Delta k}{\sqrt{\Delta k^{2}-4g}}\approx 1+\frac{2g^{2}}{%
\Delta k^{2}},
\end{equation}
and consequently to
\begin{eqnarray}
v_{a}\left( t\right) &=&e^{-i\omega _{m}t}\left\{ e^{-k_{+}t}v_{a}+\frac{ig}{%
\Delta k}\left( e^{-k_{-}t}-e^{-k_{+}t}\right) v_{b}\right\} ,  \nonumber \\
v_{b}\left( t\right) &=&e^{-i\omega _{m}t}\left\{ e^{-k_{-}t}v_{b}+\frac{ig}{%
\Delta k}\left( e^{-k_{-}t}-e^{-k_{+}t}\right) v_{a}\right\} ,
\end{eqnarray}
where
\begin{eqnarray}
k_{+} &=&k_{aa}+\frac{g^{2}}{\Delta k},  \nonumber \\
k_{-} &=&k_{bb}-\frac{g^{2}}{\Delta k}.
\end{eqnarray}
The natural decay rate for cavity $A$ ($B$) is approximately $k_{aa}$ ($%
k_{bb}$). Under the assumption we made ($k_{bb}\gg k_{aa}$), if they were
uncoupled cavity $B$ mode would decay much faster than cavity $A$. The
effect of the coupling is to alter the decay rate of both cavities, leaving
in cavity $B$ a component (which can be made large by controlling $v_{a}$)
which will decay with approximately the rate of cavity $A$. This result
resembles the classical situation of pendulum synchronization, in which the
center of mass motion is superdamped and the worst pendulum becomes phase
locked to the best one.

\section{Superposition states: controlling coherence}

\label{Schcat}We next model a feasible experimental situation where one
constructs entangled states between two cavities \cite{Davidovich2}. We
consider an initially pure entangled state
\begin{equation}
\left| \psi \left( 0\right) \right\rangle =N^{\frac{1}{2}}\left( \left|
w,0\right\rangle -\left| 0,w\right\rangle \right) ,  \label{psi}
\end{equation}
where $w$ is an index for coherent state, $N=\left( 2-2\exp \left( -\left|
w\right| ^{2}\right) \right) ^{-1}$ and clearly the density operator is
\[
\mathbf{\rho }\left( 0\right) =\left| \psi \left( 0\right) \right\rangle
\left\langle \psi \left( 0\right) \right| .
\]
The time evolution here is given by
\begin{eqnarray}
\mathbf{\rho }\left( t\right)  &=&N\left\{ \left( \left| \sigma _{1}\left(
t\right) ,\epsilon _{2}\left( t\right) \right\rangle -e^{i\phi }\left|
\epsilon _{1}\left( t\right) ,\sigma _{2}\left( t\right) \right\rangle
\right) \left( \left\langle \sigma _{1}\left( t\right) ,\epsilon _{2}\left(
t\right) \right| -e^{-i\phi }\left\langle \epsilon _{1}\left( t\right)
,\sigma _{2}\left( t\right) \right| \right) -\right.   \nonumber \\
&&\left. \left[ \left( \frac{\left| \left\langle 0|w\right\rangle \right|
^{2}-\left| \left\langle \epsilon _{1}\left( t\right) |\sigma _{1}\left(
t\right) \right\rangle \left\langle \epsilon _{2}\left( t\right) |\sigma
_{2}\left( t\right) \right\rangle \right| e^{-i\phi }}{\left\langle \epsilon
_{1}\left( t\right) |\sigma _{1}\left( t\right) \right\rangle \left\langle
\epsilon _{2}\left( t\right) |\sigma _{2}\left( t\right) \right\rangle }%
\right) \left| \sigma _{1}\left( t\right) \right\rangle \left\langle
\epsilon _{1}\left( t\right) \right| \otimes \left| \epsilon _{2}\left(
t\right) \right\rangle \left\langle \sigma _{2}\left( t\right) \right| +%
{\mathrm{h.c.}}\right] \right\} ,  \label{schstate}
\end{eqnarray}
where 
\begin{eqnarray}
\sigma _{1}\left( t\right)  &=&e^{-k_{m}t}e^{-i\omega _{m}t}f_{1}\left(
t\right) w,  \nonumber \\
\sigma _{2}\left( t\right)  &=&e^{-k_{m}t}e^{-i\omega _{m}t}f_{2}\left(
t\right) w,  \nonumber \\
\epsilon _{1}\left( t\right)  &=&e^{-k_{m}t}e^{-i\omega _{m}t}l_{1}\left(
t\right) w,  \nonumber \\
\epsilon _{2}\left( t\right)  &=&e^{-k_{m}t}e^{-i\omega _{m}t}l_{2}\left(
t\right) w
\end{eqnarray}
are also indices for coherent states. Note that the dynamics of the
non-diagonal elements is dictated by the difference between the
``distances'' between the components of the initial state, $\left|
\left\langle 0|w\right\rangle \right| ^{2}$, and a similar quantity for the
``time evolved'' states in both cavities. The condition for coherence
preservation is that 
\begin{equation}
\left| \left\langle 0|w\right\rangle \right| ^{2}=\left\langle \epsilon
_{1}\left( t\right) |\sigma _{1}\left( t\right) \right\rangle \left\langle
\sigma _{2}\left( t\right) |\epsilon _{2}\left( t\right) \right\rangle ,
\end{equation}
\textit{i.e.}: there is time evolution, but it is such that the scalar
product of the evolved states is preserved. An extreme case would be unitary
evolution, but we enforce that we only need this ``unitary-like behavior''
with respect to the vector states evolved in the initial state (\ref{psi}).

A simple way to study state's purity is its linear entropy 
\begin{equation}
\delta \left( \mathbf{\rho }\right) =Tr\left\{ \rho -\rho ^{2}\right\} .
\end{equation}
For the state of eq.\ (\ref{schstate}) one gets 
\begin{equation}
\delta \left( t\right) =\frac{\left( y^{2}\left( t\right) -1\right) \left(
x^{2}-y^{2}\left( t\right) \right) }{2y^{2}\left( t\right) \left( 1-x\right)
^{2}},
\end{equation}
where $x=\left| \left\langle 0|w\right\rangle \right| ^{2}$ and $y\left(
t\right) =\left| \left\langle \epsilon _{1}\left( t\right) |\sigma
_{1}\left( t\right) \right\rangle \right| \left| \left\langle \sigma
_{2}\left( t\right) |\epsilon _{2}\left( t\right) \right\rangle \right| $.
This latter quantity generalizes the well known result that the time scale
of coherence loss in a single cavity depends on the ``distance'' of the
states which constitute the initial superposition. Time dependent
``distances'' such as $\left| \left\langle \epsilon _{i}\left( t\right)
|\sigma _{i}\left( t\right) \right\rangle \right| ^{2}$ appear in the
analytical expression for $\delta \left( t\right) $. Note that $\delta
\left( t\right) =0$ can be obtained by setting $k_{aa}=k_{bb}=k_{ab}=k_{ba}$%
, $\omega _{a}=\omega _{b}$ and $\Delta _{aa}=\Delta _{bb}=\Delta
_{ab}=\Delta _{ba}=0$.

\begin{table}[p]
\begin{tabular}{|l|l|l|l|l|l|l|}
\hline
& $\mathbf{a}^{\dagger }\mathbf{a}\bullet $ & $\bullet \mathbf{a}^{\dagger }%
\mathbf{a}$ & $\mathbf{a}\bullet \mathbf{a}^{\dagger }$ & $\mathbf{b}%
^{\dagger }\mathbf{b}\bullet $ & $\bullet \mathbf{b}^{\dagger }\mathbf{b}$ & 
$\mathbf{b}\bullet \mathbf{b}^{\dagger }$ \\ \hline
$\mathbf{a}^{\dagger }\mathbf{a}\bullet $ & $0$ & $0$ & $-\mathbf{a}\bullet 
\mathbf{a}^{\dagger }$ & $0$ & $0$ & $0$ \\ \hline
$\bullet \mathbf{a}^{\dagger }\mathbf{a}$ & $0$ & $0$ & $-\mathbf{a}\bullet 
\mathbf{a}^{\dagger }$ & $0$ & $0$ & $0$ \\ \hline
$\mathbf{a}\bullet \mathbf{a}^{\dagger }$ & $\mathbf{a}\bullet \mathbf{a}%
^{\dagger }$ & $\mathbf{a}\bullet \mathbf{a}^{\dagger }$ & $0$ & $0$ & $0$ & 
$0$ \\ \hline
$\mathbf{b}^{\dagger }\mathbf{b}\bullet $ & $0$ & $0$ & $0$ & $0$ & $0$ & $-%
\mathbf{b}\bullet \mathbf{b}^{\dagger }$ \\ \hline
$\bullet \mathbf{b}^{\dagger }\mathbf{b}$ & $0$ & $0$ & $0$ & $0$ & $0$ & $-%
\mathbf{b}\bullet \mathbf{b}^{\dagger }$ \\ \hline
$\mathbf{b}\bullet \mathbf{b}^{\dagger }$ & $0$ & $0$ & $0$ & $\mathbf{b}%
\bullet \mathbf{b}^{\dagger }$ & $\mathbf{b}\bullet \mathbf{b}^{\dagger }$ & 
$0$ \\ \hline
$\mathbf{a}^{\dagger }\mathbf{b}\bullet $ & $-\mathbf{a}^{\dagger }\mathbf{b}%
\bullet $ & $0$ & $-\mathbf{b}\bullet \mathbf{a}^{\dagger } $ & $\mathbf{a}%
^{\dagger }\mathbf{b}\bullet $ & $0$ & $0$ \\ \hline
$\bullet \mathbf{a}^{\dagger }\mathbf{b}$ & $0$ & $\bullet \mathbf{a}%
^{\dagger }\mathbf{b}$ & $0$ & $0$ & $-\bullet \mathbf{a}^{\dagger }\mathbf{b%
}$ & $-\mathbf{b}\bullet \mathbf{a}^{\dagger }$ \\ \hline
$\mathbf{b}\bullet \mathbf{a}^{\dagger }$ & $0$ & $\mathbf{b}\bullet \mathbf{%
a}^{\dagger }$ & $0$ & $\mathbf{b}\bullet \mathbf{a}^{\dagger }$ & $-\bullet 
\mathbf{a}^{\dagger }\mathbf{b}$ & $0$ \\ \hline
$\mathbf{b}^{\dagger }\mathbf{a}\bullet $ & $\mathbf{b}^{\dagger }\mathbf{a}%
\bullet $ & $0$ & $0$ & $-\mathbf{b}^{\dagger }\mathbf{a}\bullet $ & $0$ & $-%
\mathbf{a}\bullet \mathbf{b}^{\dagger }$ \\ \hline
$\bullet \mathbf{b}^{\dagger }\mathbf{a}$ & $0$ & $-\bullet \mathbf{b}%
^{\dagger }\mathbf{a}$ & $-\mathbf{a}\bullet \mathbf{b}^{\dagger }$ & $0$ & $%
\bullet \mathbf{b}^{\dagger }\mathbf{a}$ & $0$ \\ \hline
$\mathbf{a}\bullet \mathbf{b}^{\dagger }$ & $\mathbf{a}\bullet \mathbf{b}%
^{\dagger }$ & $0$ & $0$ & $0$ & $\mathbf{a}\bullet \mathbf{b}^{\dagger }$ & 
$0$ \\ \hline
\end{tabular}
\newline
\par
\begin{tabular}{|l|l|l|l|l|l|l|}
\hline
& $\mathbf{a}^{\dagger }\mathbf{b}\bullet $ & $\bullet \mathbf{a}^{\dagger }%
\mathbf{b}$ & $\mathbf{b}\bullet \mathbf{a}^{\dagger }$ & $\mathbf{b}%
^{\dagger }\mathbf{a}\bullet $ & $\bullet \mathbf{b}^{\dagger }\mathbf{a}$ & 
$\mathbf{a}\bullet \mathbf{b}^{\dagger }$ \\ \hline
$\mathbf{a}^{\dagger }\mathbf{a}\bullet $ & $\mathbf{a}^{\dagger }\mathbf{b}%
\bullet $ & $0$ & $0$ & $-\mathbf{b}^{\dagger }\mathbf{a}\bullet $ & $0$ & $-%
\mathbf{a}\bullet \mathbf{b}^{\dagger }$ \\ \hline
$\bullet \mathbf{a}^{\dagger }\mathbf{a}$ & $0$ & $-\bullet \mathbf{a}%
^{\dagger }\mathbf{b}$ & $-\mathbf{b}\bullet \mathbf{a}^{\dagger }$ & $0$ & $%
\bullet \mathbf{b}^{\dagger }\mathbf{a}$ & $0$ \\ \hline
$\mathbf{a}\bullet \mathbf{a}^{\dagger }$ & $\mathbf{b}\bullet \mathbf{a}%
^{\dagger }$ & $0$ & $0$ & $0$ & $\mathbf{a}\bullet \mathbf{b}^{\dagger } $
& $0$ \\ \hline
$\mathbf{b}^{\dagger }\mathbf{b}\bullet $ & $-\mathbf{a}^{\dagger }\mathbf{b}%
\bullet $ & $0$ & $-\mathbf{b}\bullet \mathbf{a}^{\dagger }$ & $\mathbf{b}%
^{\dagger }\mathbf{a}\bullet $ & $0$ & $0$ \\ \hline
$\bullet \mathbf{b}^{\dagger }\mathbf{b}$ & $0$ & $\bullet \mathbf{a}%
^{\dagger }\mathbf{b}$ & $0$ & $0$ & $-\bullet \mathbf{b}^{\dagger }\mathbf{a%
}$ & $-\mathbf{a}\bullet \mathbf{b}^{\dagger }$ \\ \hline
$\mathbf{b}\bullet \mathbf{b}^{\dagger }$ & $0$ & $\mathbf{b}\bullet \mathbf{%
a}^{\dagger }$ & $0$ & $\mathbf{a}\bullet \mathbf{b}^{\dagger }$ & $0$ & $0$
\\ \hline
$\mathbf{a}^{\dagger }\mathbf{b}\bullet $ & $0$ & $0$ & $0$ & $\mathbf{a}%
^{\dagger }\mathbf{a}\bullet -\mathbf{b}^{\dagger }\mathbf{b}\bullet $ & $0$
& $-\mathbf{b}\bullet \mathbf{b}^{\dagger }$ \\ \hline
$\bullet \mathbf{a}^{\dagger }\mathbf{b}$ & $0$ & $0$ & $0$ & $0$ & $\bullet 
\mathbf{b}^{\dagger }\mathbf{b}-\bullet \mathbf{a}^{\dagger }\mathbf{a}$ & $-%
\mathbf{a}\bullet \mathbf{a}^{\dagger }$ \\ \hline
$\mathbf{b}\bullet \mathbf{a}^{\dagger }$ & $0$ & $0$ & $0$ & $\mathbf{a}%
\bullet \mathbf{a}^{\dagger }$ & $\mathbf{b}\bullet \mathbf{b}^{\dagger }$ & 
$0$ \\ \hline
$\mathbf{b}^{\dagger }\mathbf{a}\bullet $ & $\mathbf{b}^{\dagger }\mathbf{b}%
\bullet -\mathbf{a}^{\dagger }\mathbf{a}\bullet $ & $0 $ & $-\mathbf{a}%
\bullet \mathbf{a}^{\dagger }$ & $0$ & $0$ & $0$ \\ \hline
$\bullet \mathbf{b}^{\dagger }\mathbf{a}$ & $0$ & $\bullet \mathbf{a}%
^{\dagger }\mathbf{a}-\bullet \mathbf{b}^{\dagger }\mathbf{b} $ & $-\mathbf{b%
}\bullet \mathbf{b}^{\dagger }$ & $0$ & $0$ & $0$ \\ \hline
$\mathbf{a}\bullet \mathbf{b}^{\dagger }$ & $\mathbf{b}\bullet \mathbf{b}%
^{\dagger }$ & $\mathbf{a}\bullet \mathbf{a}^{\dagger } $ & $0$ & $0$ & $0$
& $0$ \\ \hline
\end{tabular}
\caption{Commutation relations $\left[ \protect\chi _{i},\protect\chi _{j}%
\right] $ for the bosonic superoperators ($\protect\chi _{i}$ and $\protect%
\chi _{j}$ are given by the line and the column respectively).}
\end{table}

\section{Experimental (im)possibilities}

\label{Exp}

It is clear that any ``unexpected'' result in the problem of two modes
interacting with one reservoir comes from the cross terms $k_{ab}$, $k_{ba}$%
, $\Delta _{ab}$ and $\Delta _{ba}$. In fact, as usual in quantum mechanics,
one is dealing with an interference phenomena. In ordinary two ways
interference phenomena (\textit{e.g.}: double slit), in order to achieve
large visibility, one tries to balance the interferometer and to guarantee
coherence between the two ways. Here, the interference is due to dissipation
rates. Balance means that each reservoir mode interacts with both
oscillators with the same strength. In a two ways interferometer, coherence
means a fixed phase relation between the two ways. The same situation
applies to coupling constants here: complete ``coherence'' (we will use
``coherence'' whenever we use the word coherence with this meaning; there
must be no confusion with the term decoherence) is achieved only when the
phase of $\frac{\alpha _{k}}{\beta _{k}}$ is independent of $k$. When the
modes are directly coupled ($g\neq 0$), another effect can also originate
cross terms, associated \textit{e.g.} to variations of the density of modes
between the two normal modes frequencies \cite{Ponte1}.

We begin by considering two field modes of distinct cavities. In principle,
the reservoir contains modes of all frequencies. It is not bad to consider
that the large wavelength modes interact with both cavities' modes in a
``coherent'' way (something similar to the dipole approximation \cite{Cohen1}%
). However, the modes close to resonance are potentially the most important
ones, as is emphasized by eq. (\ref{k_ij+delta_ij}). If on top of that the
most important reservoir modes are also field modes, this means that the
most important wavelengths are close to the wavelength of the cavities'
modes. This suggests difficulties to implement this situation with distant
cavities in the mode wavelength scale. Finally, the short wavelength modes
of the reservoir, even without very large contributions to individual decay
processes, generally will tend to destroy this ``coherence'' condition. As
long as no directly coupling is imposed, one doesn't need to consider other
effects.

The above argumentation suggests the use of two modes of one and the same
cavity. Now, if one thinks in terms of wavelengths, the most important
reservoir modes may possibly interact with the cavity modes in a
``coherent'' way. However, two (almost) degenerated modes form a two level
system, and one can think about them as two polarizations of one spatial
mode. The ``coherent'' interaction with the reservoir then defines a
superdamped and a subdamped polarizations, in a way similar to super and
subradiant modes of an $n$ atoms maser \cite{Foldi1}, and in the limit
situation, the subdamped mode would become a decoherence (and dissipation)
free subspace (or mode). But what if we change our description from the
previous polarizations to the super and subdamped ones? Then we would be
describing a system where only one polarization is coupled to the reservoir,
while the other is (almost) free. This is not the case, in general, but it
is the case for some systems. One good example is the experiment of ref. 
\cite{Kielpinski1}, where DFS principle is addressed. A classical analog is
given by two coupled harmonic oscillators, where center of mass mode usually
couples strength with the environment and decays (much) faster then the
relative motion. Another favorable situation is the dipole situation \cite
{Akram1,Ficek1} where one has a first principles model for the coupling
constants, and ``coherence'' can be naturally achieved manipulating the
dipoles.

Obviously an explicit mechanism regarding the geometry of the cavities is
still missing. It will be introduced and compared to experiment in a
forthcoming publication \cite{Magalhaes1}.

In short, we studied the effects of correlations among coupling constants of
each of two modes with a common reservoir, showing that they can smoothly
vary from two independent damped systems, to one superdamped and one free
modes, in case of perfect correlations and/or ubiquitous environment density
of modes. Examples of state evolution were given and some experimental
considerations were made, and we believe that some light was shed on
decoherence and dissipation processes for composed systems in cavity QED
when the proposed decoherence mechanism seems to work well \cite{Haroche1}.

\end{document}